\documentclass[preprintnumbers]{revtex4}

\usepackage{graphicx}
\def\mpl{\ifmmode \overline M_{Pl}\else $\overline M_{Pl}$\fi}
\def\to{\rightarrow}

\begin{document}
\bibliographystyle{revtex}

\preprint{SLAC-PUB-8960/
          P3-02}

\title{Signals For Extra Dimensions at the VLHC}

\author{Thomas G. Rizzo}

\email[]{rizzo@slac.stanford.edu}
\affiliation{Stanford Linear Accelerator Center, 
Stanford University, Stanford, California 94309 USA}

\date{\today}

\begin{abstract}
A brief overview of the signatures for several different models with extra 
dimensions at the stage II, $\sqrt s=175-200$ TeV VLHC is presented. 
In all cases the search reaches for these models in the Drell-Yan channel are 
found to be in the range of 15-80 TeV.
\end{abstract}

\maketitle

\section{Introduction}

There are now many models with extra spatial dimensions that predict the 
appearance of new physics signatures at colliders that can probe energy 
scales in 
excess of $\sim$1 to a few TeV. These models generally fall into three distinct 
classes which lead to very different phenomenologies and collider 
signatures: ($i$) those based on the large extra dimensions scenario of 
Arkani-Hamed, Dvali and Dimopoulos(ADD){\cite {add}}, which predicts the 
emission and exchange of large Kaluza-Klein(KK) towers of gravitons that are 
finely-spaced in mass; 
($ii$) those with TeV-scale dimensions(TeV){\cite {anton}, which predict the 
existence of KK excitations of the SM gauge (and possibly other) fields at 
the TeV scale and ($iii$) those with warped extra dimensions, such as the 
Randall-Sundrum Model(RS){\cite {rs}}, which predict graviton resonances with 
both weak scale masses and couplings to matter. The stage II VLHC with a 
center of mass energy $\sqrt s \sim 175-200$ TeV and a integrated luminosity 
in the range of 200-1000 $fb^{-1}$ will be able to search for and/or 
make detailed studies of models in all 
three classes. For most models of type ($i$) or ($iii$) which deal with the 
hierarchy problem, 
if no signal is observed by the time the mass 
scales probed by the VLHC are reached, the motivation behind these particular 
models will be greatly weakened if not entirely removed. 
In what follows, for simplicity and in order to avoid 
potentially difficult detector issues associated with such high energies and 
luminosities, we will focus on searches involving only the simple Drell-Yan 
process $pp \to e^+e^- +X$. From studies performed for the Tevatron and LHC 
we know that this channel provides an excellent probe of the parameter 
spaces of extra-dimensional models and we expect that this will 
continue to be true at even higher energies.

\section{Signatures}

In the SM, the Drell-Yan reaction is a result of photons and $Z$'s mediating 
the sub-process $q\bar q \to e^+e^-$. In the ADD model, graviton 
towers can also be exchanged and an additional sub-process 
$gg\to e^+e^-$, mediated solely by gravitons, also contributes{\cite {pheno}} 
at the same order. 
When summed, the effect of the graviton towers can be described through a set 
of dimension-8 operators in the limit that the center of mass energy of the 
collision process lies sufficiently 
below the cut-off scale, $M_s$, which is of order the 
size of the Planck scale in the extra dimensional space. 
In the convention used by Hewett{\cite {pheno}}, the 
contribution of 
spin-2 tower exchange can be expressed in terms of the scale, $M_s$, and 
a sign, $\lambda$. In the Drell-Yan case, the deviations of the cross section 
are found to be essentially $\lambda$-independent but with a strong $M_s$ 
dependence. Current experimental constraints from LEP and the 
Tevatron{\cite {greg}} tell us that $M_s \geq 1$ TeV, and values for $M_s$ as 
large as the low 10's of TeV may be conceivable in this framework. 
The distortion of the Drell-Yan cross section at large lepton pair invariant 
masses at the VLHC through these dimension-8 operators can trivially probe such 
high mass scales in a manner similar to searches for compositeness-type contact 
interactions. The shape of the deviation from the SM with varying dilepton 
mass will tell us that the underlying physics arises 
due to dimension-8 operators; in addition at large masses 
the dilepton angular distribution would conform to the shape expected due to 
the dominance of spin-2 exchange thus nailing down the gravitational origin 
of these effects. 
These types of deviations can be easily seen in Fig.~\ref {p3-02_fig1} for two 
different center of mass energies and integrated luminosities. It is clear 
from these results that there are enough statistics available in the high 
mass dilepton data to provide sensitivity to values of $M_s$ in excess of 
25-40 TeV at the VLHC. Of course, in the ADD model, we would expect to see 
new physics in other channels as well, in particular the existence of monojets 
from graviton emission.

\begin{figure}[htbp]
\centerline{
\includegraphics[width=5.5cm,angle=90]{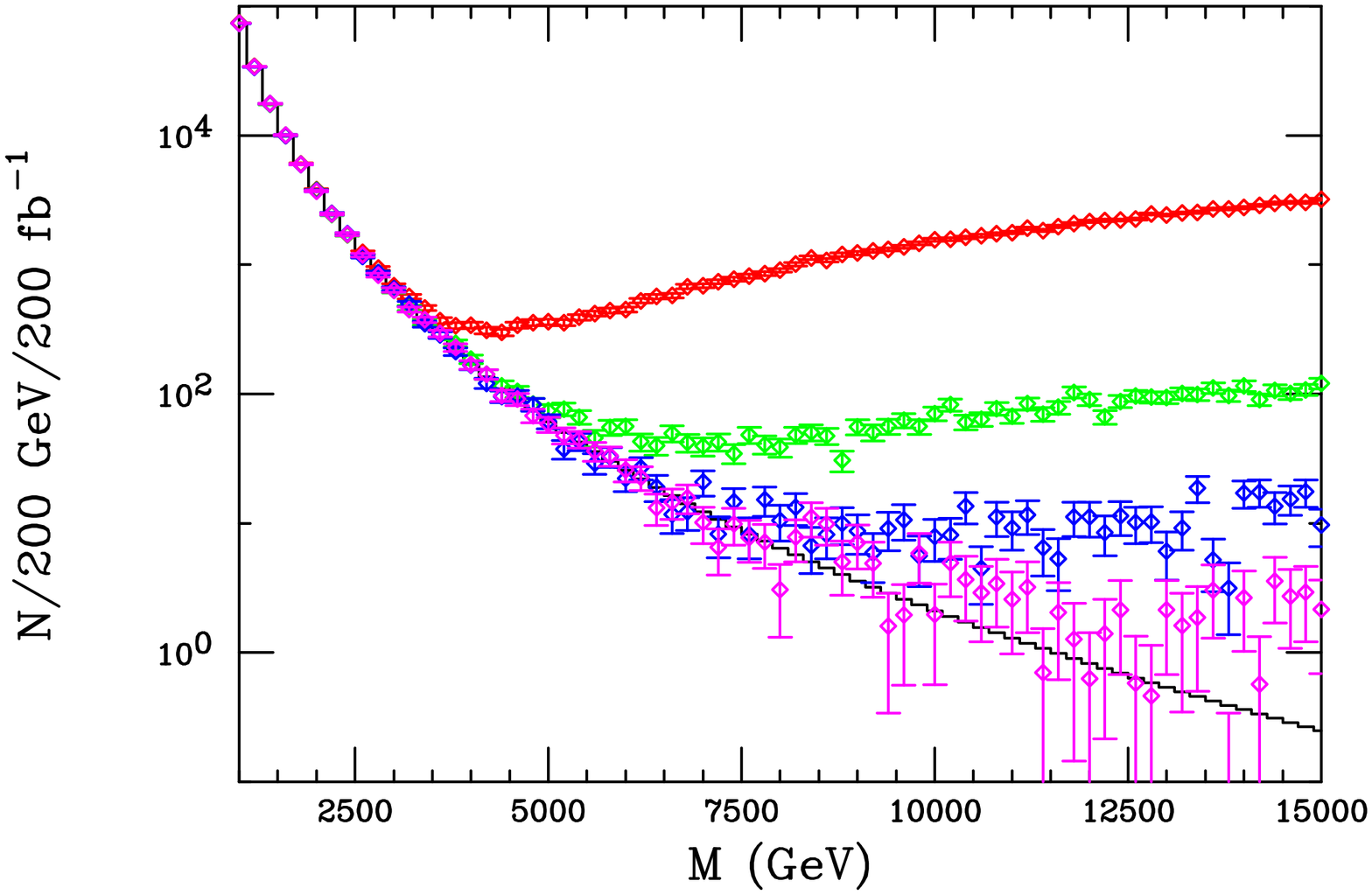}
\hspace*{5mm}
\includegraphics[width=5.5cm,angle=90]{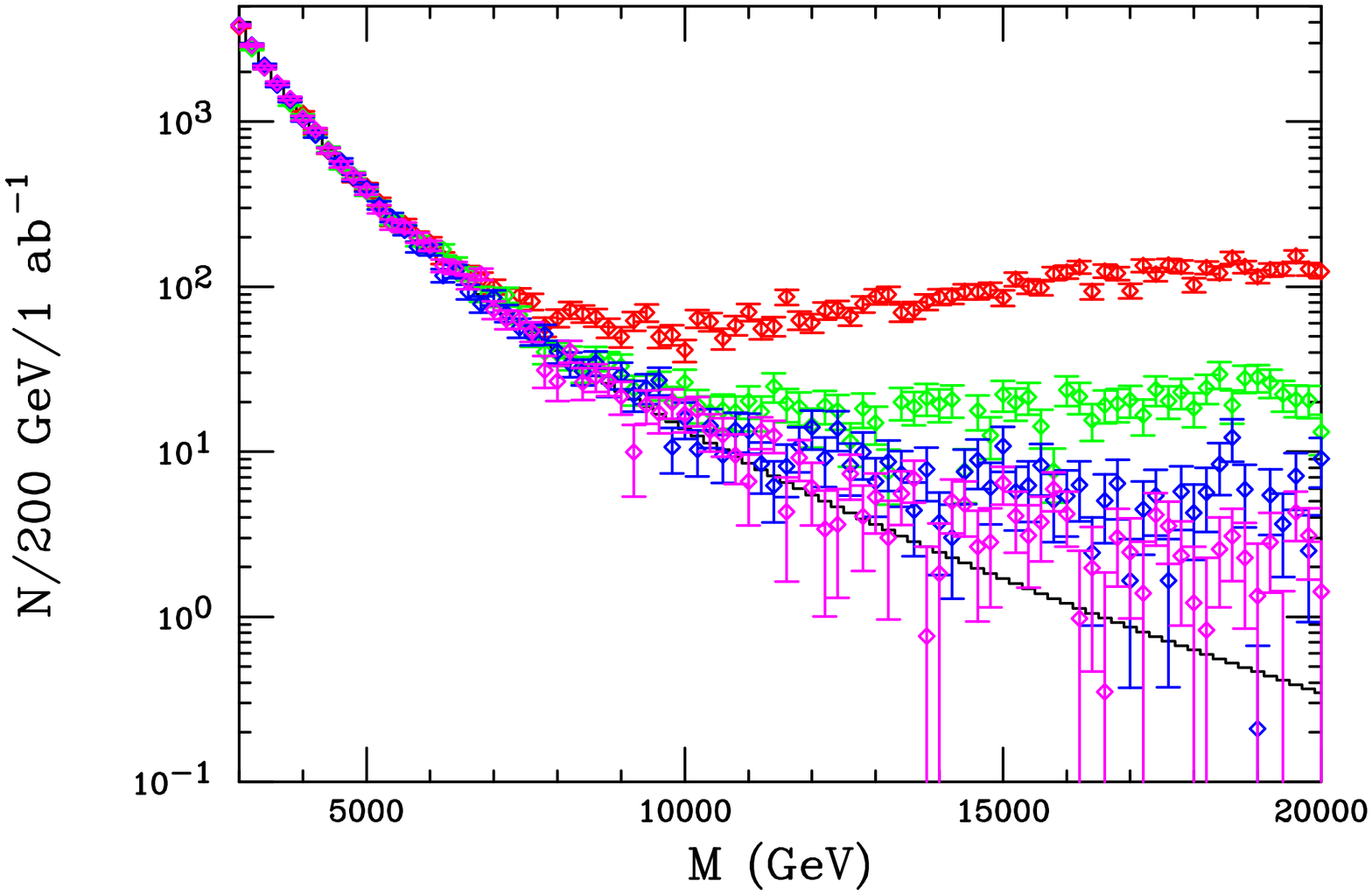}}
\vspace*{0.1cm}
\caption{Event rate per 200 GeV mass bin for the Drell-Yan process as a 
function of the lepton pair invariant mass at a 
$\sqrt s=175(200)$ stage II VLHC in the left(right) panel for different 
integrated luminosities. A rapidity cut $\eta_l<2.5$ on both leptons has been 
applied. The solid histogram is the SM background in both 
cases whereas the `data' points are the predictions of the ADD model. In the 
left(right) panel the red, green, blue and magenta points correspond to the 
assumption that $M_s=10,15,20$ or 25$(20,25,30$ or 35) TeV, respectively.}
\label{p3-02_fig1}
\end{figure}

In the simplest versions of TeV scale theories with extra dimensions, only the 
SM gauge fields are in the bulk whereas the fermions remain at the orbifold 
fixed points; Higgs fields may lie at the fixed points or propagate in the 
bulk. Of course, more complicated scenarios with very different phenomenology 
are possible. In such a simple case with only one extra dimension 
it has been shown that the current high precision 
electroweak data can place a lower bound on the mass of the first KK excited 
gauge boson in excess 
of $\simeq$ 4 TeV{\cite {bunch}}. In such a model, to a good 
approximation, the masses of the KK tower states are given by $M_n=nM_c$, 
where $M_c$ is the compactification scale. 
For this one extra dimensional example all of the excited KK states 
have identical couplings to the SM fermions. In this case, while the first 
KK state may be observable at the LHC the higher KK modes will not be due to 
their rather small cross sections arising from their large masses 
and a higher energy machine will be necessary 
to explore the KK spectrum. Fortunately the VLHC does provide a window into 
such high scales as can be seen from Fig.~\ref{p3-02_fig2}.
Here one observes that for `low' values of the compactification scale, 
$M_c \leq 
10$ TeV, as many as 5 to 6 KK excitations will be easily observable and that 
values of $M_c$ as large as $\sim 50-60$ TeV will be directly probed in 
Drell-Yan by the stage II VLHC.

\begin{figure}[htbp]
\centerline{
\includegraphics[width=5.5cm,angle=90]{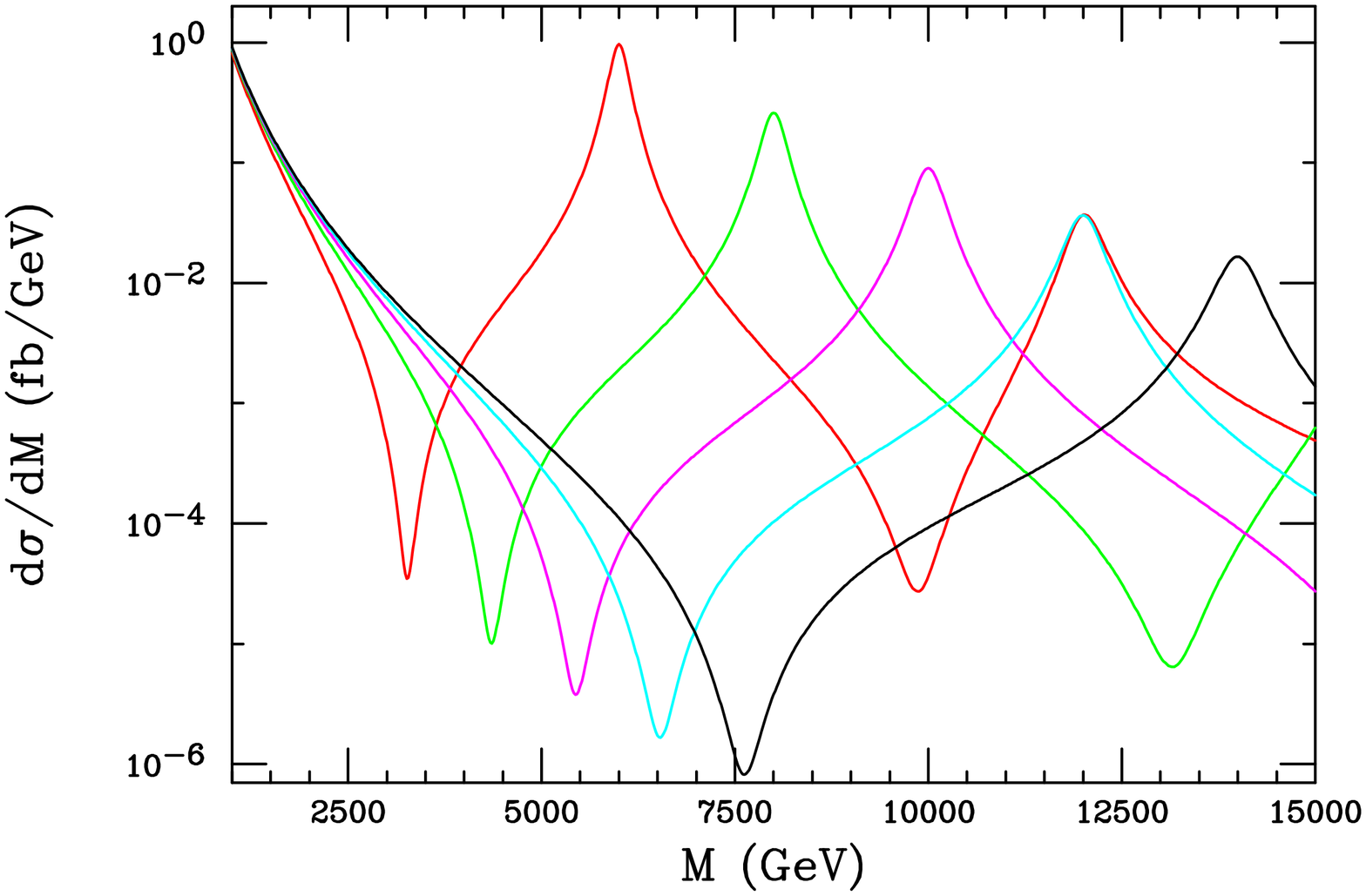}
\hspace*{5mm}
\includegraphics[width=5.5cm,angle=90]{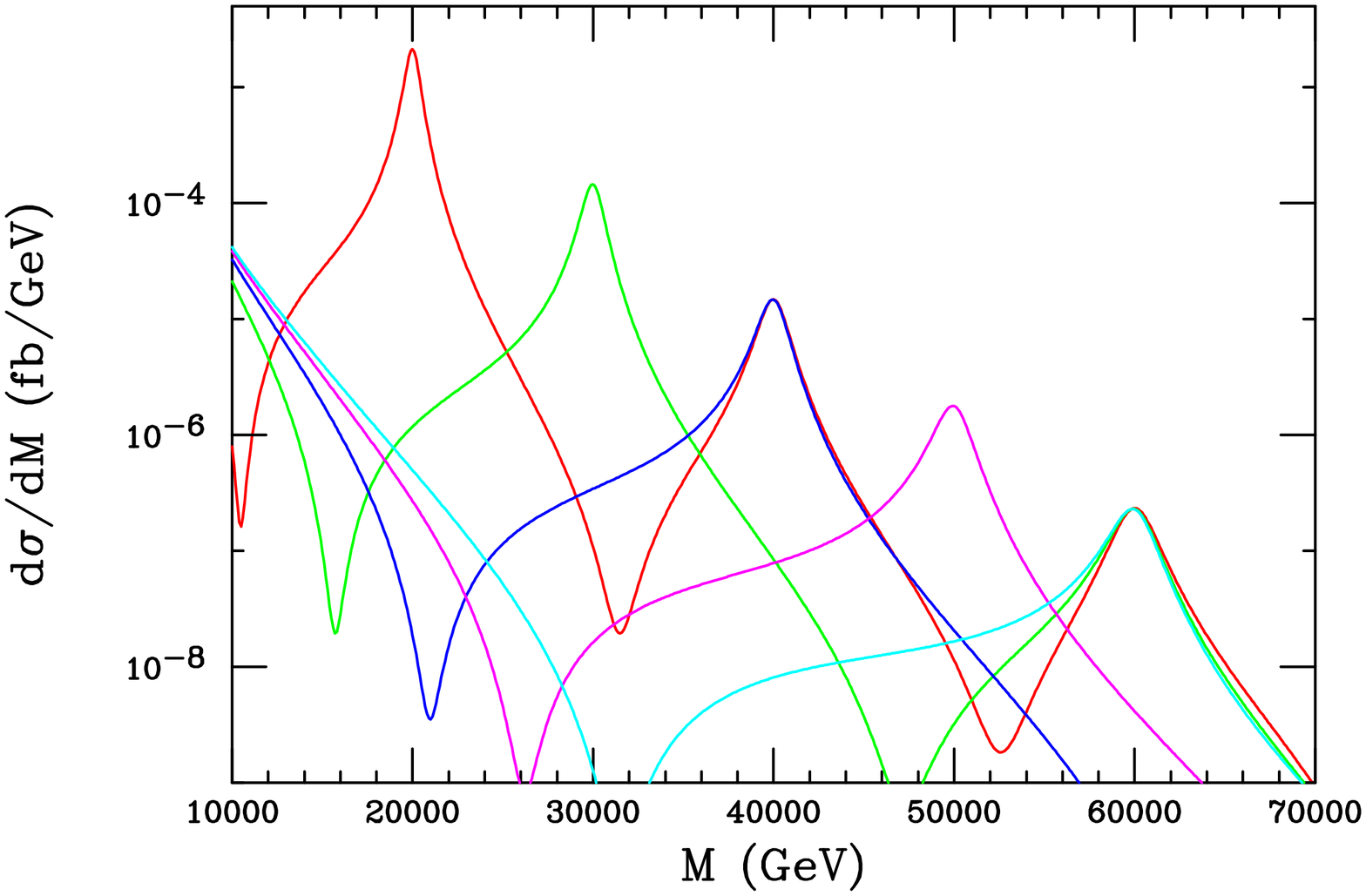}}
\vspace*{0.1cm}
\caption{Spectra for $\gamma/Z$ KK excitations in Drell-Yan 
assuming only one extra dimension and all SM fermions sitting at the same 
orbifold fixed point. In the left(right) panel the 
subsequent curves(from left to right) 
correspond to compactification scales of 6,8,10, 
12 and 14 (20,30,40,50 and 60) TeV, respectively. $\sqrt s=175$ TeV and a 
rapidity cut as above have been assumed.}
\label{p3-02_fig2}
\end{figure}

 The VLHC may be even more useful if the number of extra dimensions is 
greater than or equal to two; in this case, still keeping the fermions at the 
orbifold fixed points, the bounds from precision data are expected to be 
stricter than in the one-dimensional case but are less quantitatively 
precise since the naive evaluation of the relevant 
sums over KK states are divergent. (Hence the exact limits are sensitive 
to the physics that cuts off these sums.) Also one finds that 
the masses and couplings of 
KK excitations become both level and compactification-scheme dependent thus 
leading to a rather complex KK spectrum in Drell-Yan. In fact it 
will be necessary to observe a rather large part of the lower portion of the 
spectrum in order to experimentally 
determine the number of extra dimensions and how they are compactified. 
Fortunately the VLHC may allow for such a detailed study provided $M_c$ is not 
too large.

\begin{figure}[htbp]
\centerline{
\includegraphics[width=5.5cm,angle=90]{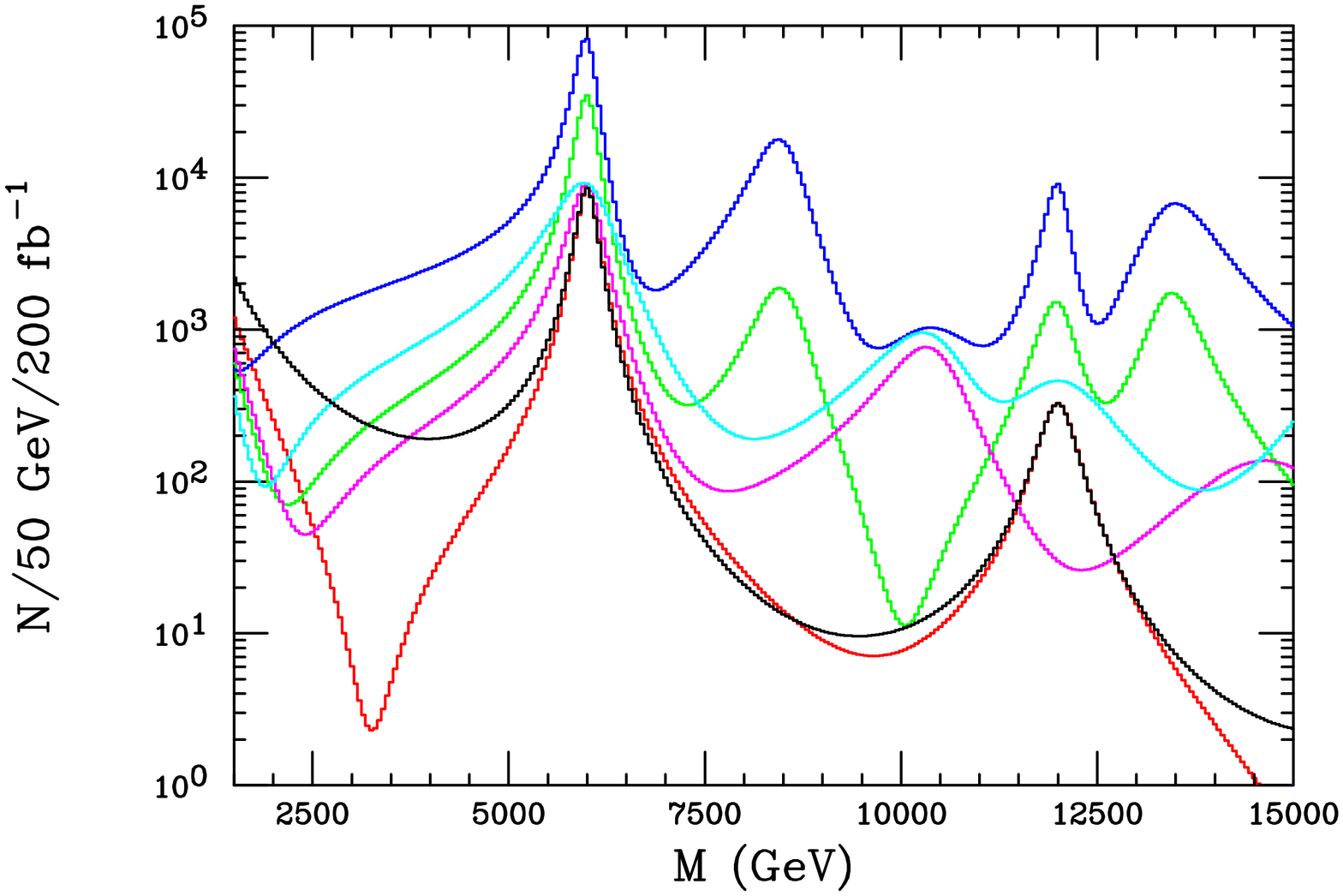}
\hspace*{5mm}
\includegraphics[width=5.5cm,angle=90]{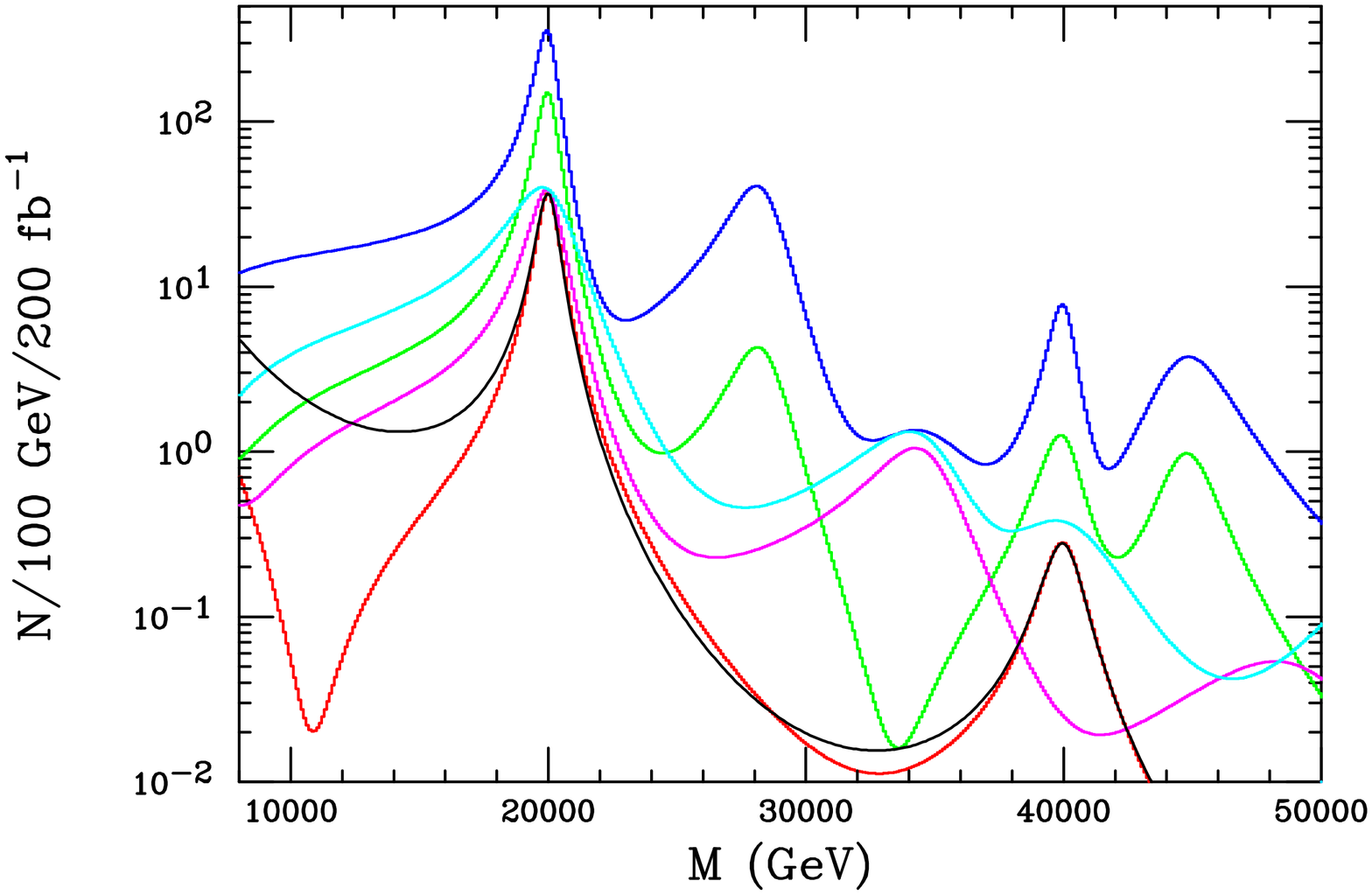}}
\vspace*{0.1cm}
\caption{Binned excitation spectra in Drell-Yan for several models with one 
or more 
extra dimensions at the VLHC for $\sqrt s=175$ TeV and with the same rapidity 
cut as above. In the 
left(right) panel $M_c=6(20)$ TeV has been assumed.}
\label{p3-02_fig3}
\end{figure}

Some sample KK excitation spectra for 
a number of different TeV-scale models of this kind with more than one extra 
dimension are shown in Fig.~\ref{p3-02_fig3}. Note that these various spectra 
are quite distinctive. Specifically, the 
measurements of the locations of the peaks and their relative heights and 
widths can be used to uniquely identify a given extra-dimensional model. 
From this figure it is clear that the VLHC can be used to differentiate the 
many possible models even for large compactification scales $\simeq 20$ TeV or 
higher through detailed cross section measurements.

\begin{figure}[htbp]
\centerline{
\includegraphics[width=5.5cm,angle=90]{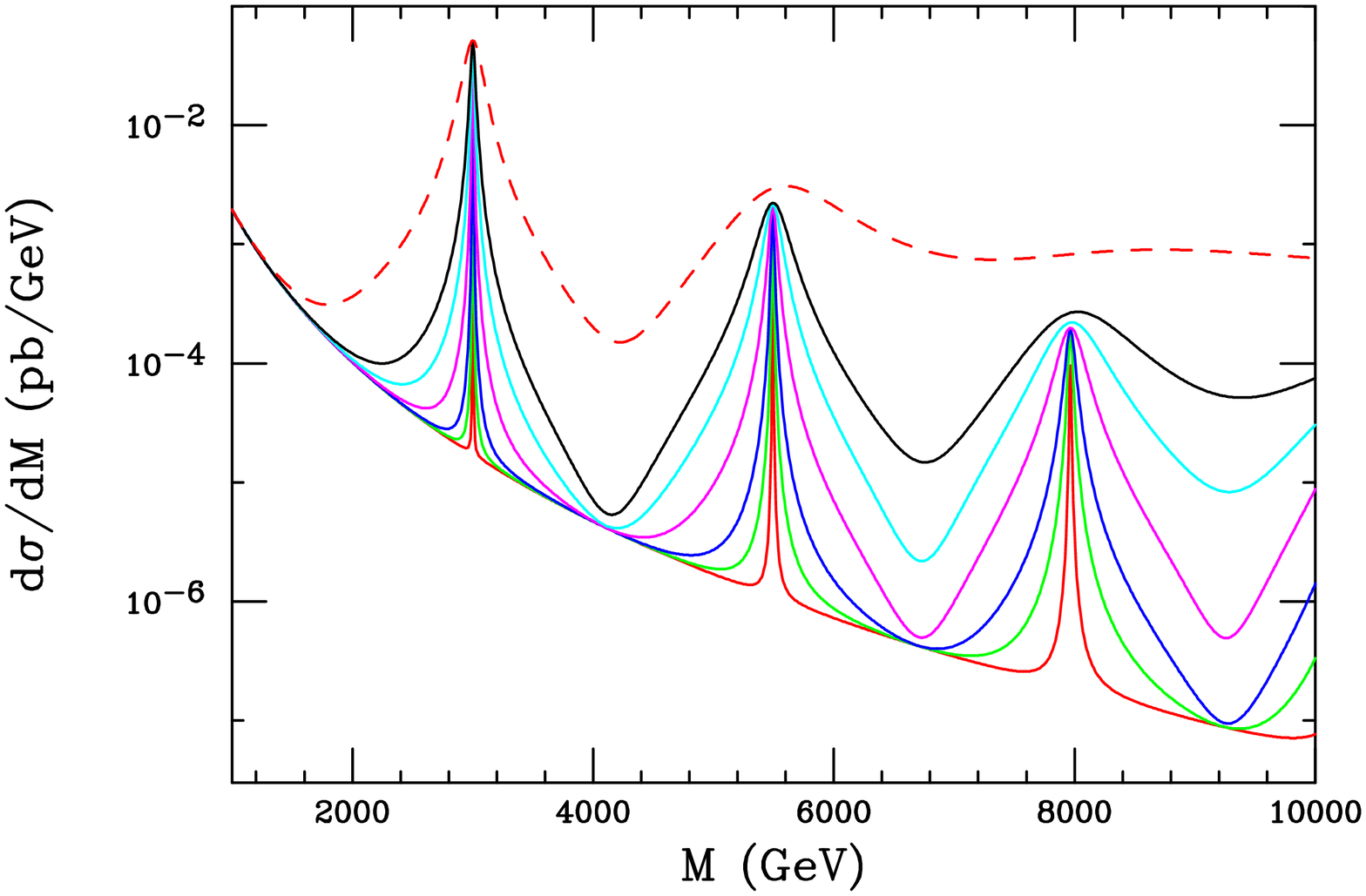}
\hspace*{5mm}
\includegraphics[width=5.5cm,angle=90]{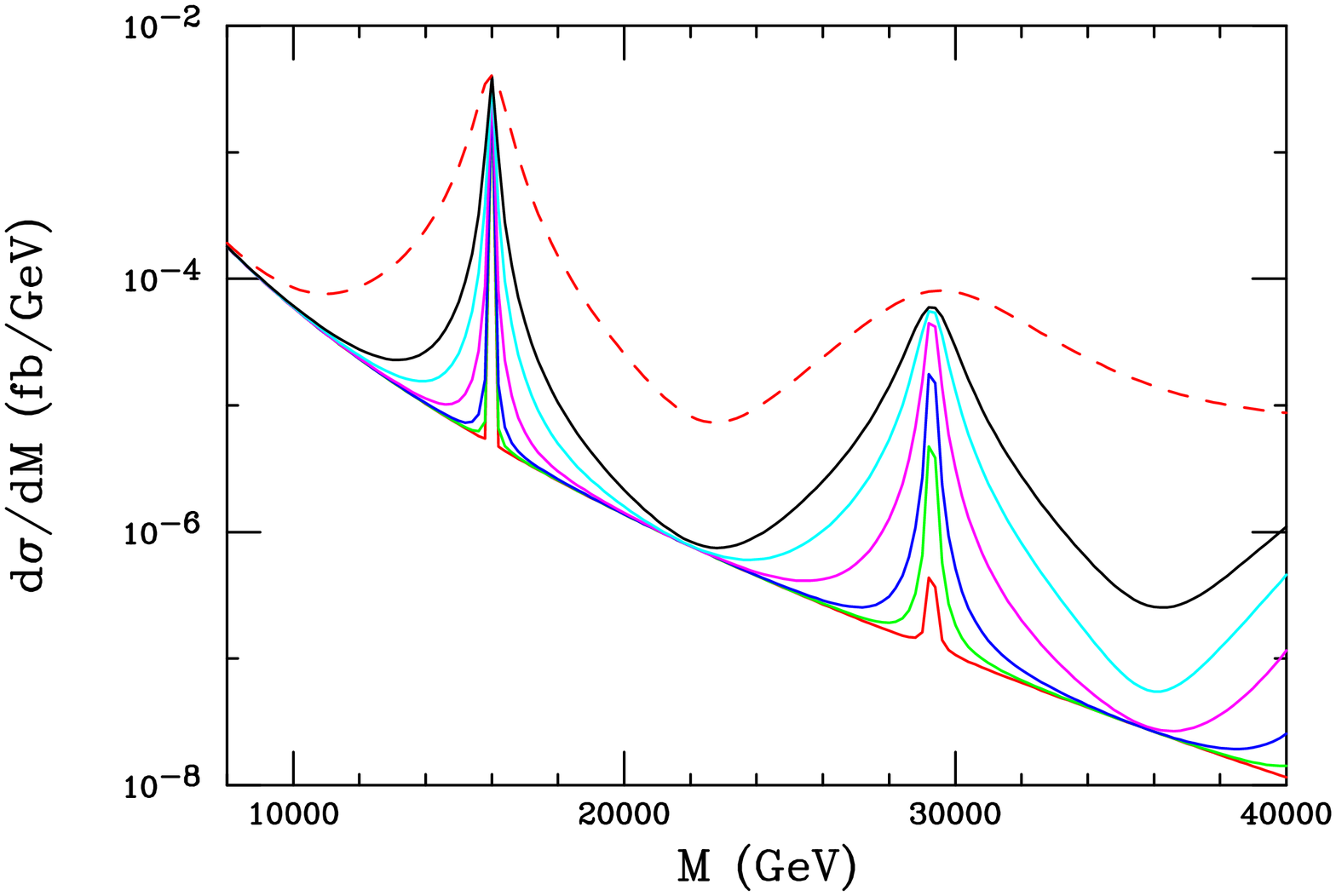}}
\vspace*{0.1cm}
\caption{Excitation of KK gravitons in the RS model in Drell-Yan collisions 
at the $\sqrt s=175$ TeV VLHC. From most narrow to widest the curves correspond 
to values of the parameter $c=k/\mpl$ of 0.01,0.02,0.03,0.05,0.075,0.1 and 
0.2, respectively. The left(right) panel assumes that the lightest KK graviton 
excitation has a mass of 3(16) TeV.}
\label{p3-02_fig4}
\end{figure}

The last case we consider is the RS model wherein, as discussed above, we 
expect to produce TeV-scale graviton resonances in many channels{\cite {dhr}} 
including Drell-Yan. In its simplest version, with only one extra dimension, 
two distinct branes, 
and with all of the SM fields remaining on the TeV-brane, this model has only 
two fundamental parameters: the mass of the first KK state (from which all 
the others can be determined) and an additional 
parameter, $c=k/\mpl$, which we expect to be smaller than but not too far 
from unity. This parameter essentially controls the effective coupling 
strength of the gravitons(when expressed in terms of the mass of the lowest 
lying KK state) and thus also the widths of the corresponding resonances in 
Drell-Yan. This implies that the overall production cross 
section is highly $c$-dependent which can be seen explicitly in 
Fig.~\ref{p3-02_fig4}. Here we find that a 200 TeV VLHC with 1~$ab^{-1}$ of 
integrated luminosity will be able to observe the first RS KK 
excitation for masses as large as 15-30 TeV for values of $c$ in the 
interesting range $\sim$0.01-0.2. By studying the lepton pair mass bins near 
the resonance, the decay angular distribution may be determined. This would 
demonstrate that a spin-2 particle is being produced while measurements of 
the relative branching fractions to other clean decay modes, such as 
$\gamma\gamma$, can prove that we are producing gravitons. 
We also see that for a lighter KK spectrum, the 
VLHC will observe and determine the masses of a reasonable number of KK 
resonances. Using the ratios of these KK masses one would be able to 
demonstrate that we had discovered the states as predicted by the 
five-dimensional 
RS model since their masses are in the ratios of the roots of 
the $J_1$ Bessel function. With it's high luminosity it also seems possible 
that the VLHC will be able to 
perform a detailed study of some of the more exotic decays of the heavier 
graviton states{\cite {us}} that may occur in this model; to examine this 
possibility will require more elaborate simulations.

\section{Discussion and Conclusion}

From the discussion above it is clear that the mass reach in Drell-Yan at 
the stage II VLHC offers an excellent window on many different models with 
extra dimensions. In some cases the lack of observation of a positive 
signature at such mass scales would serious weaken the justification for these 
models.

%
\def\MPL #1 #2 #3 {Mod. Phys. Lett. {\bf#1},\ #2 (#3)}
\def\NPB #1 #2 #3 {Nucl. Phys. {\bf#1},\ #2 (#3)}
\def\PLB #1 #2 #3 {Phys. Lett. {\bf#1},\ #2 (#3)}
\def\PR #1 #2 #3 {Phys. Rep. {\bf#1},\ #2 (#3)}
\def\PRD #1 #2 #3 {Phys. Rev. {\bf#1},\ #2 (#3)}
\def\PRL #1 #2 #3 {Phys. Rev. Lett. {\bf#1},\ #2 (#3)}
\def\RMP #1 #2 #3 {Rev. Mod. Phys. {\bf#1},\ #2 (#3)}
\def\NIM #1 #2 #3 {Nuc. Inst. Meth. {\bf#1},\ #2 (#3)}
\def\ZPC #1 #2 #3 {Z. Phys. {\bf#1},\ #2 (#3)}
\def\EJPC #1 #2 #3 {E. Phys. J. {\bf#1},\ #2 (#3)}
\def\IJMP #1 #2 #3 {Int. J. Mod. Phys. {\bf#1},\ #2 (#3)}
\def\JHEP #1 #2 #3 {J. High En. Phys. {\bf#1},\ #2 (#3)}

\end{document}